\documentclass[twocolumn,showpacs,preprintnumbers,amsmath,amssymb]{revtex4}
\input psfig.sty

\usepackage{graphicx}
\usepackage{dcolumn}
\usepackage{bm}

\begin{document}

\title{Gravitational lensing constraints on dark energy from modified Friedmann equations}

\author{Abha Dev} \email{abha@ducos.ernet.in}
\affiliation{Department of Physics and Astrophysics, University of Delhi, Delhi - 110007, India
}

\author{J. S. Alcaniz} \email{alcaniz@dfte.ufrn.br}

\affiliation{Astronomy Department, University of Washington, Seattle,
Washington, 98195-1580, USA}

\affiliation{Departamento de Fisica, Universidade Federal do Rio Grande do Norte, C.P. 1641, Natal - RN,
59072-970, Brazil}

\author{Deepak Jain} \email{deepak@ducos.ernet.in}
\affiliation{Deen Dayal Upadhyaya College, University of Delhi, Delhi - 110015, India
}

\date{\today}

\begin{abstract}

The existence of a dark energy component has usually been invoked as the most plausible way to explain the
recent observational results. However, it is also well known that effects arising from new physics (e.g.,
extra dimensions) can mimic the gravitational effects of a dark energy through a modification of the Friedmann
equation. In this paper we investigate some observational consequences of a flat,
matter dominated and accelerating/decelerating scenario in which this modification is given by $H^{2} =
g(\rho_m, n, q)$ where $g(\rho_m, n, q)$ is a new function of the energy density $\rho_m$, the so-called
generalized Cardassian
models. We mainly focus our attention on the constraints from statistical properties of gravitationally lensed
quasars on the parameters $n$ and $q$ that fully characterize the models.
We show that these models are in agreement with the current gravitationally lensed quasar data for a large
interval of the $q - n$ parametric space. The dependence of the acceleration
redshift (the redshift at which the universe begins to accelerate) on these parameters is also briefly
discussed.

\end{abstract}

\maketitle

\section{Introduction}

Over the last years, a considerable number of high quality observational data have transformed radically the
field of cosmology. Results from distance measurements of type Ia supernovae (SNe Ia) \cite{perlmutter}
combined with Cosmic Microwave Background (CMB) observations \cite{bern}
and dynamical estimates of the quantity of matter in the universe \cite{calb} seem to indicate that the simple
picture provided by the standard cold dark matter scenario (SCDM) is not enough.
These observations are usually explained by introducing a new hypothetical energy component with negative
pressure, the so-called dark energy or {\emph{quintessence}} \cite{peebles}. If confirmed, the existence of
this dark component would also provide a definitive piece of information connecting the inflationary flatness
prediction with astronomical data.

On the other hand, it is also well known that not less exotic mechanisms like, e.g., geometrical effects from
extra
dimensions may be capable of explaning such observational results. The
basic idea behind these ``braneworld cosmologies" is that our 4-dimensional Universe would be a surface or a
brane embedded into a higher
dimensional bulk space-time to which gravity could spread \cite{rand}. In some of these scenarios the observed
acceleration
of the Universe can be explained (without dark energy) from the fact that the bulk gravity sees its own
curvature term on the brane acting as a negative-pressure dark component which accelerates the Universe
\cite{dvali}. A natural conclusion from these and other similar studies is that dark energy, or rather, the
gravitational
effects of a dark energy could actually be achieved from a modification of the Friedmann equation arising
from new physics.

Following this reasoning, several authors have recently
investigated cosmologies with a modified Friedmann equation from
extra dimensions as an alternative explanation for the recent
observational data. For example, Sahni \& Shtanov \cite{sahni}
investigated a new class of braneworld models which admit a wider
range of possibilities for dark energy than do the usual
quintessence scenarios. For a subclass of the parameter values,
they showed that the acceleration of the universe in this class of
models can be a transient phenomena which could help reconcile an
accelerating universe with the requirements of string/M-theory
\cite{fis}. Recently, Dvali \& Turner \cite{turner} explored the
phenomenology and detectability of a correction on the Friedmann
equation of the form $(1 - \Omega_m)H^{\alpha}/H_o^{\alpha - 2}$,
where $\Omega_m$ is the matter density parameter, $H$ is the
Hubble parameter (the subscript ``o" refers to present time) and
$\alpha$ is a parameter to be adjusted by the observational data.
Such a correction behaves like a dark energy with an effective
equation of state given by $\omega = -1 + \alpha/2$ in the recent
past and like a cosmological constant ($\omega = -1$) in the
distant future. Also based on extra dimensions physics, Freese \&
Lewis \cite{freese} proposed the so-called {\emph{Cardassian
expansion}}, a model in which the universe is flat, matter
dominated and currently accelerated. In the Cardassian universe,
the new Friedmann equation is given by $H^{2} = g(\rho_m)$, where
$g(\rho_m)$ is an arbitrary function of the matter energy density
$\rho_m$. The first version of these scenarios had $g(\rho_m) =
A\rho_m + B\rho_m^{n}$, with the second term driving the
acceleration of the universe at a late epoch after it becomes
dominant (a detailed discussion for the origin of this Cardassian
term from extra dimensions physics can be found in
\cite{freese01}). Although being completely different from the
physical viewpoint, it was promptly realized that by identifying
some free parameters Cardassian models and quintessence scenarios
parameterized by an equation of state $p = \omega \rho$ predict
the same observational effects in what concerns tests involving
only the evolution of the Hubble parameter with the redshift
\cite{freese}. More recently, several forms for the function
$g(\rho_m)$ have been proposed \cite{freese1}. In particular, Wang
{\it et al.} \cite{wang} have studied some observational
characteristics of a direct generalization of the original
Cardassian model. According to these authors, the observational
expressions in this new scenario are very different from generic
quintessence cosmologies and fully determined by two dimensionless
parameters, $n$ and $q$. Observational constraints from a variety
of astronomical data have been also investigated recently, both in
the original Cardassian model \cite{zong} and in its generalized
versions \cite{gat}. Perhaps the most interesting feature of these
models is that although being matter dominated, they may be
accelerating and can still reconcile the indications for a flat
universe ($\Omega_{\rm{total}} = 1$) from CMB observations with
the clustering estimates that point consistently to $\Omega_m
\simeq 0.3$ with no need to invoke either a new dark component or
a curvature term. In these scenarios, it happens through a
redefinition of the value of the critical density
\cite{freese,wang} (see also below).

The aim of this paper is to explore some observational
consequences of these generalized Cardassian (GC) scenarios. We
mainly focus our attention on the constraints that can be placed
on the free parameters of the model ($n$ and $q$) from statistical
properties of gravitationally lensed quasars. We show that for a
large interval of these parameters, this class of models are fully
compatible with the current gravitationally lensed quasar data. We
also investigate the dependence of other observational quantities
like the deceleration parameter, age-redshift relation and the
acceleration redshift on the parameters $n$ and $q$.

This paper is organized in the following way. In Sec. II the field
equations and distance formulas are presented. We also derive the
expression for the deceleration parameter, age of the Universe and
discuss the redshift at which the accelerated expansion begins. We
then proceed to analyze the constraints from lensing statistics
and on these scenarios in Sec. III. We end the paper by
summarizing the main results in the conclusion Section.

\section{The model: basic equations}

The modified Friedmann equation for GC models is \cite{wang}
\begin{eqnarray}
H^{2} =  \frac{8\pi G \rho_m}{3}\left[1 + \left(\frac{\rho_{card}}{\rho_m}\right)^{q(1 - n)}\right]^{1/q},
\end{eqnarray}
where $n$ and $q$ are two free parameters to be adjusted by the observational data, $\rho_{{card}} =
\rho_o(1 + z_{{card}})^{3}$ is the energy density at which the two terms inside the bracket are equal and
$\rho_o$
is the present day matter density. Note that for values of $n = 0$ and $q = 1$, GC scenarios reduce to Cold
Dark Matter models with a cosmological constant ($\Lambda$CDM). As explained in \cite{wang}, the first term
inside the bracket dominates
initially in such a way that the standard picture of the early universe is completely maintained. At a
redshift $z_{{card}}$, these two terms become equal and, afterwards, the second term
dominates, leading or not to an accelerated expansion (note, however, that $z_{card}$ is not necessarily the
acceleration
redshift. See, for instance, the discussion below on the deceleration parameter).

Evaluating Eq. (1) for present day quantities we find
\begin{equation}
H_o^{2} = \frac{8\pi G \rho_o}{3}\left[1 + (1 + z_{card})^{3q(1 - n)}\right]^{1/q}.
\end{equation}
From this equation, we see that $\rho_o$ is also the critical density that now can be written as
\begin{equation}
\rho_o = \rho_{c,old} \times \left[1 + (1 + z_{card})^{3q(1 - n)}\right]^{-{1/q}},
\end{equation}
where $\rho_{c,old} = 1.88 \times 10^{-29}h^{2}\rm{gm/cm^{3}}$ is the standard critical density and $h$
is the present day Hubble parameter in units of 100 ${\rm{km.s^{-1}Mpc^{-1}}}$. Note that for some
combinations of the parameters $z_{card}$, $n$ and $q$ the critical density can be much lower than the one
previously estimated. In other words it means that in the context of GC models it is possible to make the
dynamical estimates of the quantity of matter that consistently point to $\rho_o \simeq
(0.2-0.4)\rho_{c,old}$
compatible with the observational evidence for a flat universe from CMB observations and the inflationary
flatness prediction with no need of a dark energy component (see \cite{freese} for a more detailed
discussion). In Fig. 1 we show a generalized version of the Figure 1 of \cite{freese} in which the plane
$z_{card} - n$ is displayed for selected values of $q$. The contours are labeled indicating the fraction of
the standard critical density for different combinations of $z_{card}$ and $n$. In particular, the points
inside the shadowed area delimited by the contours 0.2 - 0.4 are roughly consistent with the present
clustering estimates \cite{calb}.

From Eq. (3), we see that the observed matter density parameter in GC models can be written as
\begin{equation}
\Omega_m^{obs} = \frac{\rho_o}{\rho_{c,old}} = \left[1 + (1 + z_{card})^{3q(1 - n)}\right]^{-{1/q}},
\end{equation}
which, from now on, we will fix at 0.3 (in accordance to dynamical estimates of $\rho_m$ \cite{calb}) in order
to discuss
the acceleration redshift below and the lensing versus redshift test in the next section.

\begin{figure}
\centerline{\psfig{figure=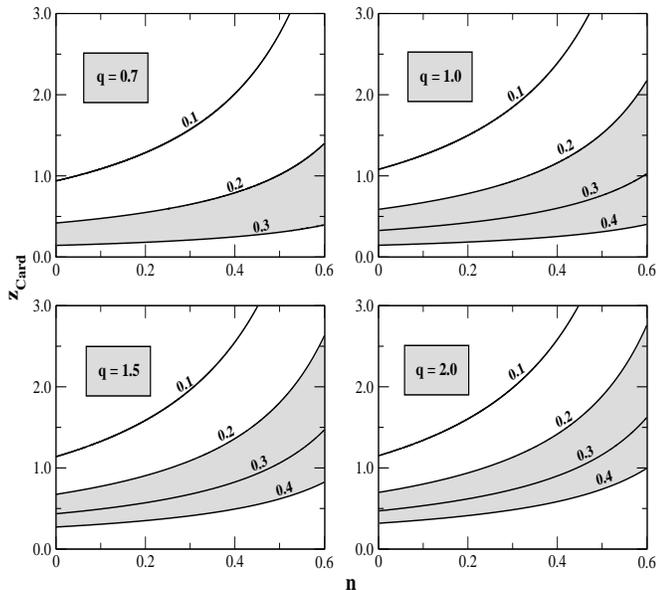,width=3.5truein,height=3.3truein,angle=-90}
\hskip 0.1in}
\caption{$z_{card} - n$ diagrams for the ratio $\rho_o/\rho_{c,old}$ (Eq. 4) and selected values of $q$. The
contours are labeled indicating the corresponding fraction of the standard critical density.  Points
inside the shadowed area are roughly consistent with the present clustering estimates \cite{calb}.}
\end{figure}

For the GC expansion parameterized by $n$ and $q$, the deceleration parameter as a function of the redshift
has the following form
\begin{equation}
q(z) \equiv - \frac{\ddot{R}R}{\dot{R}^{2}} = -1 + \frac{1}{2}\frac{d{\rm{ln}}f^{2}(z, \Omega_m^{obs}, q,
n)}{d{\rm{ln}}(1 + z)},
\end{equation}
where an overdot denotes derivative with respect to time, $R(t)$ is the cosmological scale factor and $f(z,
\Omega_m^{obs}, q, n)$ is given by Eq. (7). Figure
2 shows the behavior of the deceleration parameter as a function of redshift for selected
values of $n$ and $q$. As discussed earlier, although completely dominated by matter, GC scenarios allow
periods of accelerated expansion for some combinations of the parameters $n$ and $q$. Note that the present
acceleration is basically determined by the value of $n$ and that the smaller its value the more accelerated
is the present expansion for a given value of $q$. For example, for $q = 1.5$ and $n = 0$, GC models
accelerate presently faster ($q_o \simeq -0.75$) than flat $\Lambda$CDM scenarios with $\Omega_{\Lambda} =
0.7$ ($q_o
\simeq -0.5$) although the acceleration redshift is almost identical ($z_a \simeq 0.7$) while for the same
value of $q$
and $n = 0.5$ we find $q_o \simeq -0.13$ and $z_a \simeq 0.52$. In order to make clear the difference between
$z_{card}$ and $z_a$, for the latter values of $q$ and $n$, we find directly from Eq.(4) $z_{card} = 1.06$. It
means that although becoming dominant at $z_{card} \simeq 1$ the second term of Eq. (1) will drive an
accelerated expansion only $\sim 1.7$ Gyr later, at $z_a \simeq 0.52$.

From the above equations, it is straightforward to show that the age-redshift relation is now
given by (the total expanding age of the Universe $t_o$ is obtained by taking $z = 0$)
\begin{equation}
t_z = \frac{1}{H_o}\int_{o}^{x'} {dx \over x f(x, \Omega_m^{obs}, q, n)},
\end{equation}
where  $x' = {R(t) \over R_o} = (1 + z)^{-1}$ is a convenient integration variable and the dimensionless
function $f(x, \Omega_m^{obs}, q, n)$, obtained from Eqs. (1) and (4), is written as
\begin{eqnarray}
f(x, \Omega_m^{obs}, q, n)  =  \left\{\frac{\Omega_m^{obs}}{x^{3}}\left[1 + \frac{(\Omega_m^{obs})^{-q} -
1}{x^{-3q(1
- n)}}\right]^{1/q}\right\}^{\frac{1}{2}}.
\end{eqnarray}

\begin{figure}
\centerline{\psfig{figure=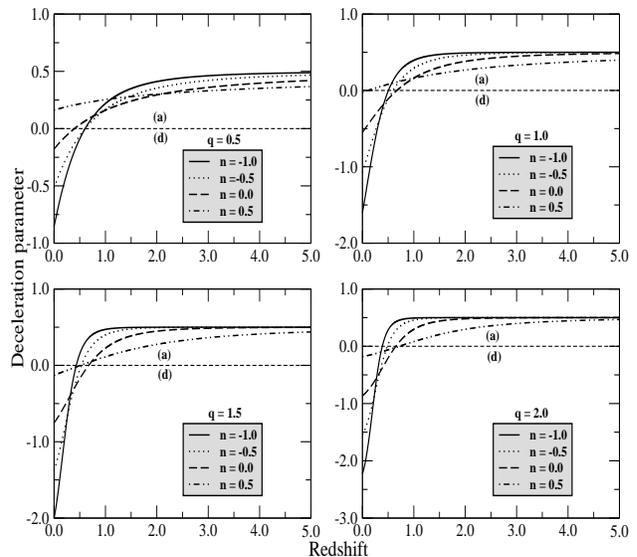,width=3.5truein,height=3.3truein,angle=-90}
\hskip 0.1in}
\caption{Deceleration parameter as a function of redshift for some selected values of
$n$ and $q$ and $\Omega_m^{obs} = 0.3$. The horizontal line labeled (a)/(d) ($q_o = 0$) divides models
with a accelerating (a) or decelerating (d) expansion at a given redshift. As discussed in the text, the
present value of $q_o$ is basically determined by the value of $n$.}
\end{figure}

The comoving distance $r_1(z)$ to a light source located at $r =
r_1$ and $t = t_1$ and observed at $r = 0$ and $t = t_o$ can be
expressed as
\begin{equation}
r_1(z) = \frac{1}{R_o H_o}\int_{x'}^{1} {dx \over x^{2}f(x, \Omega_m^{obs}, q, n)}.
\end{equation}
In order to derive the
constraints from lensing statistics in the next Section we shall deal with the concept
of angular diameter distance. For the class of GC models here investigated, the angular diameter distance,
$D_{LS}(z_L, z_S) = {R_or_1(z_L, z_S)/(1 + z_S)}$, between two objects,
for example a lens at $z_L$ and a source (galaxy) at $z_S$, reads
\begin{eqnarray}
D_{LS}(z_L, z_S) & = & \frac{H_o^{-1}}{(1 + z_S)} \times \\ \nonumber & & \times
 \int_{x'_S}^{x'_L} {dx \over x^{2} f(x, \Omega_m^{obs}, q, n)} .
\end{eqnarray}

\section{Lensing constraints}

In this Section we use statistics of gravitationally lensed quasars to place limits on the free parameters of
GC scenarios. We work
with a sample
of 867 ($z > 1$) high
luminosity optical quasars which includes 5 lensed quasars.
Our sample consists of data from the following  optical lens surveys:
HST Snapshot survey \cite{HST}, Crampton survey
\cite{Crampton}, Yee survey \cite{Yee},
Surdej survey \cite{Surdej}, NOT Survey \cite{Jaunsen}
and FKS survey \cite{FKS}. Since the main difference between the analysis performed in this Section and the
previous ones that use
gravitational lensing statistics to constrain cosmological parameters is the cosmological model that here is
being considered, we refer the reader to previous works for detailed formulas and calculational methods
(see, for instance, \cite{1CSK,lensing}). In order to perform our analysis we use the Schechter luminosity
function
with the lens
parameters for E/SO galaxies taken from
Madgwick {\it et al.} \cite{Mad}, i.e., $\phi_{\ast} = 0.27 h^{3}\,10^{-2}
{\rm{Mpc}}^{-3}$, $\alpha = -0.5$, $\gamma = 4$, $v_{\ast} = 220\,{\rm{km/s}}$ and
$F^{*} = 0.01$.

\begin{figure}
\centerline{\psfig{figure=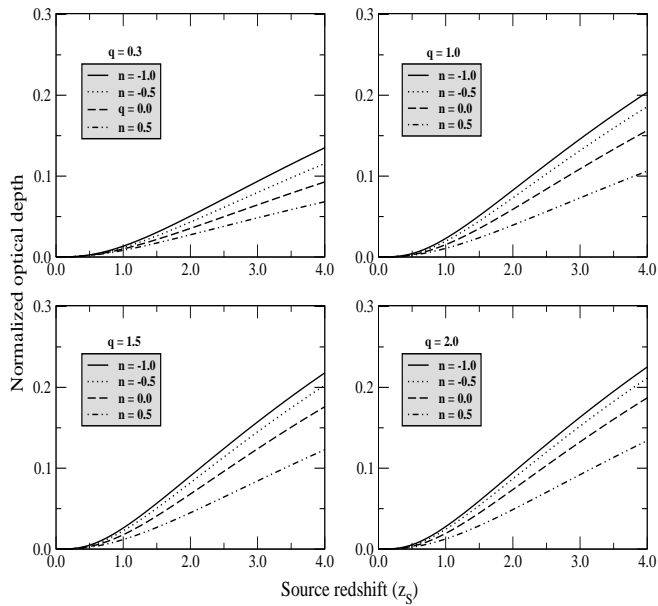,width=3.5truein,height=3.3truein,angle=-90}
\hskip 0.1in}
\caption{Normalized optical depth ($\tau/F^{*}$) as a function of the source redshift $z_S$ for some
selected  values of $n$ and $q$ and a fixed value of $\Omega_m^{obs} = 0.3$.}
\end{figure}

\begin{figure}
\centerline{\psfig{figure=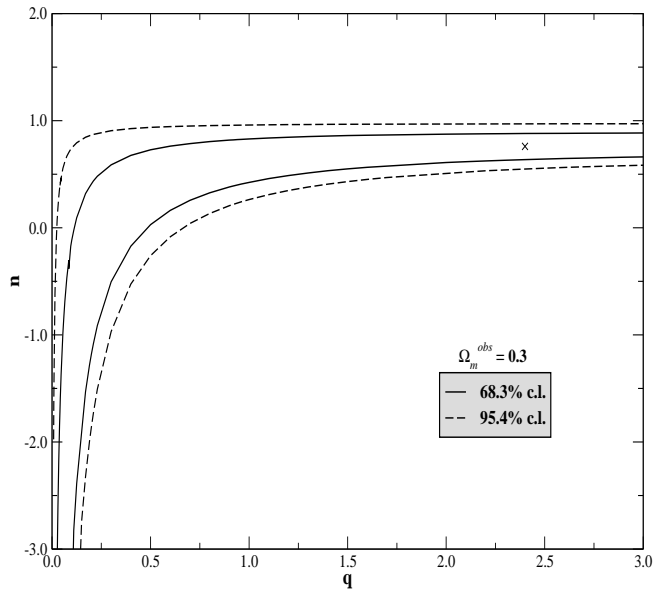,width=3.5truein,height=3.3truein,angle=-90}
\hskip 0.1in}
\caption{Confidence regions in the plane $n - q$ arising from lensing
statistics for a fixed value of $\Omega_m^{obs} = 0.3$. Solid (dashed) lines indicate contours of constant
likelihood
at $68.3\%$ c.l. ($95.4\%$ c.l.). The best-fit model is indicated by ``$\times$".}
\end{figure}

The total optical depth $\tau(z)$ along the
line of sight from an observer at $z = 0$ to a source at $z_S$ is given by
\begin{equation}
\tau(z_S) = \frac{F^{*}}{30} \left[D_{OS}(1+z_{S})\right]^{3} R_{o}^{3}.
\label{atau}
\end{equation}
where $D_{OS}$ is the angular diameter distance (Eq. 9) from the
observer to the source. In Fig. 3 for the fixed value of
$\Omega_m^{obs} = 0.3$ we show the normalized optical depth as a
function of the source redshift for values of $q = $ 0.3, 1.0, 2.0
and 3.0 and $n = $ -1.0, -0.5, 0.0 and 0.5. Note that a decrease
in the value of $n$ at fixed $\Omega_m^{obs}$ and $q$ tends to
increase the optical depth for lensing. For $q = 1.0$ at $z_S =
3.0$, the value of $\tau/F^{*}$ for $n = 0.5$ is down from that
one for $n = -1.0$ by a factor of $\sim 2.02$, while the same
values of $n$ and $q = 2.0$ provides values for $\tau/F^{*}$ that
are up from the previous ones only by a factor of $\sim$ 1.1 and
1.25, respectively. It clearly shows that the optical depth is a
more sensitive function to the parameter $n$ than to the index
$q$. As commented earlier, this particular feature is also noted
for the other analyses discussed in this paper.

The likelihood function is defined by
\begin{equation}
{\cal{L}} = \prod_{i=1}^{N_{U}}(1-p^{'}_{i})\,\prod_{k=1}^{N_{L}}
p_{k}^{'}\,p_{ck}^{'},
\label{LLF}
\end{equation}
where $N_{L}$ is the number of multiple-imaged lensed quasars, $N_{U}$ is the number of unlensed
quasars, and $p_{k}^{'}$ and $p_{ck}^{'}$ are, respectively, the probability of quasar $k$ to be lensed and
the
configuration probability (see \cite{1CSK} for details).

\begin{figure}
\centerline{\psfig{figure=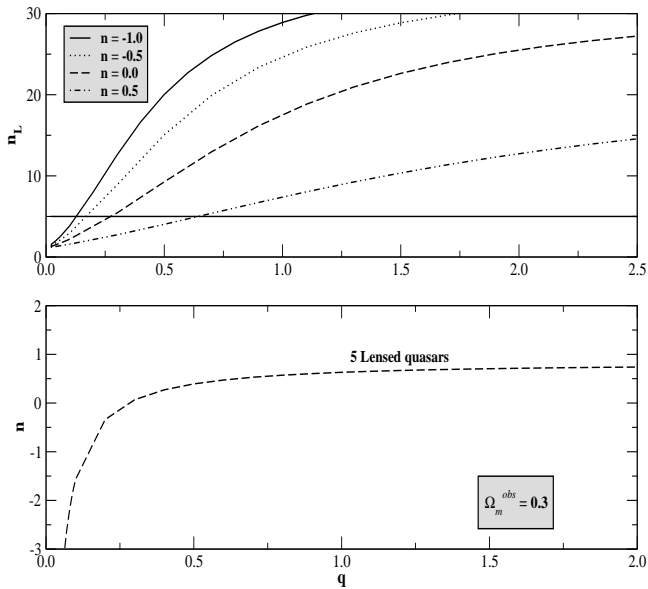,width=3.5truein,height=3.3truein,angle=-90}
\hskip 0.1in}
\caption{{\bf{a)}} Predicted number of lensed quasars as a function of the parameter $q$ for selected values
of $n$, $\Omega_{m}^{obs} = 0.3$ and image separation $\Delta \theta \leq 4^{"}$. The horizontal solid line
indicates $n_L = 5$. {\bf{b)}} Contour for five
lensed quasars in the parametric space $q - n$.}
\end{figure}

Figure 4 shows contours of constant likelihood ($68.3\%$ and
$95.4\%$) in the parameter space $n - q$. From the above equation
we find that the maximum value of the likelihood function is
located at $n = 0.76$ and $q = 2.4$ ($z_{card} \simeq 4.14$) which
corresponds to a currently decelerated universe with $q_o \simeq
0.16$, i.e., apparently in contradiction to the SNe results (it is
worth mentioning that a low-density decelerated model is not ruled
out by SNe Ia data alone \cite{mesa}, although such a model is
strongly disfavored in the light of the recent CMB data). At the
1$\sigma$ level, however, almost the entire range of $q$ (if we
consider for example $0 < q \leq 4$) is compatible with the
observational data for a fixed value of $\Omega_{m}^{obs} = 0.3$.
As observed earlier, this result suggests that a large class of GC
scenarios is in accordance with the current gravitational lensing
data. For the sake of comparison, we also note that the GC
best-fit model obtained from our analysis and the one obtained for
general quintessence scenarios with an equation of state $p_x =
\omega_x \rho_x$ (XCDM) are very alike. For example, for XCDM
models a similar analysis shows that the maximum value of the
likelihood function is located at $\Omega_{\rm{m}} = 0.0$ and
$\omega_x = -0.2$ \cite{waga} which corresponds to a decelerated
model with a deceleration parameter $q_o = 0.2$ and a total
expanding age of $8.1h^{-1}$ Gyr. The best-fit for GC model also
corresponds to a decelerated scenario with $q_o = 0.16$ and a
total age of the order of $8.7h^{-1}$ Gyr. We suspect that this
similarity may be associated with the fact that the likelihood
analysis is more sensitive to $n$ than to $q$ and, as commented
earlier, the original Cardassian scenario (which depends only on
$n$) and XCDM models predict very similar observational results
\cite{freese}. These particular values of $q$ and $n$ for the
best-fit GC model provide a predicted age of the Universe old
enough to accommodate some recent age estimates of high-$z$
objects. For example, at $z = 1.55$ and $z = 1.43$ Eq. (6)
provides for $H_o = 65$ ${\rm{kms^{-1}Mpc^{-1}}}$ $t_z = 3.76$ Gyr
and $t_z = 4.01$ Gyr, respectively, i.e., values that are in
agreement with the age estimates for the radio galaxies LBDS
53W091 and LBDS 53W069 \cite{dunlop}. For the recent discovery of
the quasar APM 08279+5255 \cite{komossa} at $z = 3.91$, however,
these values of $q$ and $n$ provide $t_z = 1.52$ Gyr while the age
estimate for this object lies between 2.0 - 3.0 Gyr (a similar
problem is also faced by the concordance $\Lambda$CDM model
\cite{jailson}).

In Fig. 5a we show the expected number of lensed quasars, $n_L = \sum\, p_{i}^{'}$ (the
summation is over a given quasar sample), as a function of the index $q$ for some selected values of $n$. As
indicated in the figure, the horizontal solid line stands for $n_L = 5$, that is the number of lensed quasars
in our sample.  By this analysis, one
finds that models with ($n$, $q$) = (-1.0, 0.13), (-0.5, 0.17), (0.0, 0.28) and (0.5, 0.65) predict exactly 5
lensed quasars. In Fig. 3b we display
the contour for five lensed quasars in the parametric space $q - n$. As a general result, this analysis shows
that a large number of models can accommodate the current gravitational lensing data.

\section{Conclusion}

The possibility of an accelerating universe from distance
measurements of type Ia supernovae constitutes one of the most
important results of modern cosmology. These observations
naturally lead to the idea of a dominant dark energy component
with negative pressure once all known types of matter with
positive pressure generate attractive forces and decelerate the
expansion of the universe. On the other hand, the realization that
dark energy or the effects of dark energy could be a manifestation
of a modification to the Friedmann equation arising from extra
dimension physics has opened up an unprecedented opportunity to
establish a more solid connection between particle physics and
cosmology. Many ``braneworld scenarios" have been proposed in the
recent literature with most of them presenting interesting
features which make them a natural alternative to the standard
model. Here we have analyzed some observational consequences of
one of these scenarios, the so-called generalized cardassian
expansion recently proposed in Ref. \cite{wang}. We have studied
the observational constraints on the parameters $n$ and $q$ that
fully characterize the model from statistical properties of
gravitational lensing. From this analysis we have found that at
$1\sigma$ level a large class of these scenarios is in agreement
with the current lensing data with the maximum of the likelihood
function located at $n = 0.76$ and $q = 2.4$ which corresponds to
a decelerated model with $q_o = 0.16$ and a predicted age of the
Universe of the order of $t_o = 8.7h^{-1}$ Gyr. Naturally, only
with a more general analysis, possibly a joint investigation
involving different classes of cosmological tests, it will be
possible to show whether or not this class of models constitutes a
viable alternative to the standard scenario.

\begin{acknowledgments}
The authors are very grateful to Zong-Hong Zhu for helpful discussions. JSA is supported by the Conselho
Nacional de Desenvolvimento
Cient\'{\i}fico e Tecnol\'{o}gico (CNPq - Brasil) and CNPq (62.0053/01-1-PADCT III/Milenio).
\end{acknowledgments}


\end{document}